\documentstyle[12pt]{article}
\begin{document}

\def\simlt{\mathrel{\lower .5ex \rlap{$\sim$}\raise .5ex \hbox{$<$}}}
\def\simgt{\mathrel{\lower .5ex \rlap{$\sim$}\raise .5ex \hbox{$>$}}}

\begin{titlepage}
\vspace*{1cm}
\begin{center}

{\Large \bf Scaling Behavior of the Activated Conductivity in a
Quantum Hall Liquid}

\vspace*{0.5cm}
{\bf S. Das Sarma} and {\bf Dongzi Liu}

\vspace*{0.5cm}
{\em Center for Superconductivity Research\\
Department of Physics\\
University of Maryland\\
College Park, Maryland 20742}

\vspace*{1cm}

\end{center}


\begin{abstract}

 We propose a scaling model for the universal longitudinal conductivity
near the mobility edge for the integer quantum Hall liquid. We fit
our model with available experimental data on exponentially activated
conductance near the Landau level tails in the integer quantum Hall
regime.
We obtain quantitative agreement between our scaling model and the
experimental data over a wide temperature and magnetic field range.

\end{abstract}
\end{titlepage}


	One of the remarkable fundamental features of the quantum Hall
effect(QHE) phenomenon[1] is its intimate connection with the
two-dimensional(2D) strong-field Landau level localization problem[2].
Our current understanding of QHE is based[1] on the following
scenario: The strong external magnetic field splits the 2D system into
Landau levels(LL) which are broadened by disorder invariably present
in the system; the electron states at the center($E_c$) of the
disorder-broadened LL  are extended whereas all the other states at LL
tails are localized with the localization length diverging at the
mobility edge $E_c$; at $T=0$ the 2D system undergoes a
metal-insulator transition at $E_c$ where $\sigma_{xy}$ jumps from one
quantum Hall plateau to the next and $\sigma_{xx}$ has a universal
LL-independent value $\sigma^c$(believed to be $\sigma^c=e^2/h$) --
$\sigma_{xx}$ is zero(at $T=0$) everywhere except at $E_F=E_c$(with
$E_F$ the Fermi level) whereas $\sigma_{xy}$ is quantized everywhere
except at $E_c$(where it jumps from one plateau to the next). At
finite temperature (or, in fact, for samples with finite sizes), the
situation has to be modified[1-3] somewhat because when the
``effective'' system size becomes comparable to the localization
length at a particular energy, the system behaves as metallic rather
than insulating, producing a small finite width of the extended state
region around $E_c$. Consequently, $\sigma_{xx}$ acquires a
temperature dependent width($\Delta B$) around $E_c$, and,
$\sigma_{xy}$ goes from one plateau to the next with a finite slope.
It is now experimentally well established[3] that the strong-field
magneto-transport properties of a 2D system show scaling  localization
behavior around this $\sigma_{xx}$(or, $\rho_{xx}$) peak where
$\sigma_{xy}$(or, $\rho_{xy}$) is changing from one plateau to the next
and the chemical potential $E_F$ is passing through $E_c$ near the
center of LL.

   Wei {\em et al}[3] reported the first experimental evidence in
support of the
universal scaling behavior of the magneto-resistivity near the
metal-insulator transition. They found that the slope of the transition
for $\rho_{xy}$ from one QH step to the next in the changing magnetic
field($d\rho_{xy}/dB$),  and
the half-width($\Delta B$) of the finite peak of $\rho_{xx}$ both
have exactly the same scaling dependence on temperature,
{\em i.e.} $d\rho_{xy}/dB
\propto T^{-\kappa }$ and $\Delta B\propto T^{\kappa}$.
Experimentally[3], it was found
$\kappa \approx 0.42$ in an InGaAs/InP heterostructure for the two lowest
Landau levels. But the extent to which $\kappa$ is really a universal
constant independent of Landau level index and sample materials is still
somewhat uncertain[2].

   While the experimental support for strong-field scaling
localization comes entirely from studying the $\rho_{xx}$ peaks in the
``metallic'' region ({\it i.e.} $E_F\approx E_c$)(and,
the associated behavior of $\rho_{xy}$), detailed experimental
results are also available in the literature on the activated
temperature dependence
of $\rho_{xx}$ minima in the ``insulating'' region
centering in the QH plateau regime of
$\rho_{xy}$.
The purpose of this paper is to analyze the $\rho_{xx}(T)$ minima
behavior from the viewpoint of the scaling theory and to show that the
existing data in the literature on the activated behavior of
$\rho_{xx}$ at LL tails is consistent with the scaling theory.

   A different kind of universal behavior of the temperature activation of the
dissipative (longitudinal) conductance $\sigma_{xx}$ was observed by
Clark {\em et al} in
the integer quantum Hall plateau regime[4]. A sample of their data is
shown in Fig.1. They found that in a certain range of temperature,
$\sigma_{xx}$ may be written as
$\sigma_{xx}=\sigma^{c}e^{-\Delta /k_{B}T}$, where $\Delta\approx
|E_F-E_c|$ is the activation energy.  By extrapolating the fit
to the linear regime of the data to $1/T=0$, it was found that the
prefactor $\sigma^{c}\simeq e^{2}/h$ is independent of sample
and the Landau level index.
There is also a detailed experimental investigation[5] of
$\sigma_{xx}$ minima by Katayama {\it et al} who studied the
activated regime as well. In this paper, we provide a scaling analysis of
these results. Our scaling model assumes that the only conduction
mechanism operating at the $\sigma_{xx}$ minima is the thermal
activation of localized electrons from the Fermi level to the nearest
mobility edge, where the usual scaling theory should apply. As such we
are neglecting any direct transport contribution ({\it e.g.} variable
range hopping) of the localized carriers themselves, which should not
scale because it happens very far from $E_c$ and is presumably outside
the critical regime. Our model, therefore, does not apply at very low
temperatures where variable range hopping dominates and scaling breaks down.

   Recently, Lee {\em et al} [6] provided a theoretical basis for Clark's
experiments[4]. They found a relationship between the zero
temperature diagonal conductivity at a quantum Hall step and the
prefactor of the activated longitudinal conductivity in the plateau
regime, which according to Ref.6, has the following form:
\begin{equation}
  \sigma_{xx}\simeq
2\sigma_{1}^{c}e^{-\Delta/k_{B}T}S_{2}(\frac{k_{B}T}{\Gamma})
\label{s2}
\end{equation}
where $S_{2}(y)$ is a universal scaling function, and $\Gamma$ is the
energy width (arising, for example, from inelastic scattering at
finite $T$) of the energy band around $E_c$
of delocalized states.
In the temperature range in which $k_{B}T<<\Delta $, but $k_{B}T$
still large enough so that variable-range hopping (which dominates at
very low temperature) can be ignored, the prefactor $\sigma^{c}_1$ in
the experiment of Ref.4 is given by $2S_{2}(0)\sigma_{1}^{c}$ which is
manifestly universal.
Thus, according to Ref.6, activated conduction in the QH step is a
rather simple universal scaling function.
Unfortunately, however,
there exist certain technical errors in this theoretical derivation
of Ref.6 which render it invalid,
and, as
a result, within the scheme of  Ref.6, the longitudinal conductivity, even
neglecting variable-range hopping, is not given[7] by an activated
form (such as Eq.(1))
with a universal prefactor, as was originally claimed in Ref.6.

    In this {\em Rapid Communication}, we propose a scaling model for
the universal activated longitudinal conductivity near
the mobility edge for the
integer quantum Hall liquid. We use the best-fit analysis to compare
our model with the experimental data.
Our conclusion is that while the experimental activated conductivity data are
clearly consistent with the universal
scaling behavior, it may not be appropriate
to determine the precise value of this universal conductance simply by
extrapolating the
experimental data to $1/T=0$. This is because the behavior of
$\sigma_{xx}$ minima at the QH steps is more complicated than a simple
Arrhenius activation behavior. On the other hand, we find that a
simple extrapolation[4] does provide the correct order of magnitude of
$\sigma^c$, which lends credence to the analysis of Ref.4.

    We first establish the relation between the activated
conductance and the critical conductance near the mobility edge.
Consider a system of non-interacting 2D electrons in a perpendicular
magnetic field in the presence of a random disorder potential. The
real part of the dc conductivity  for non-interacting electrons is
given by
\begin{equation}
  \sigma_{xx}=-\int\frac{\partial f_{FD}}{\partial E}\sigma_{1}(E)
\end{equation}
where $f_{FD}$ is the Fermi-Dirac thermal distribution function.
At zero temperature, $-\frac{\partial f_{FD}}{\partial E}=\delta
(E-E_{F})$, so $\sigma_{xx}=\sigma_{1}(E_{F})$. It is obvious from
the definition that $\sigma_{1}(E)$ is the longitudinal conductivity at
the Fermi energy $E_F$, and,  in general, at finite temperatures
$\sigma_{1}(E)$ should also be
temperature dependent.

The single
particle eigenstates in the Landau band are all localized except at
one critical energy $E_c$
in the center of the Landau band, where the
electronic state is extended for the infinitely large system at zero
temperature.  In the scaling regime the localization length $\xi $
 for the electronic
state diverges as the energy approaches the critical energy $E_{c}$,
{\em i.e.} $\xi (E) \propto |E-E_{c}|^{-\nu }$,
$\nu $ is currently believed[2] to be universal($\simeq 7/3$).

  At zero temperature, only the extended state at the critical energy
contributes to $\sigma_{xx}$, {\em i.e.}
\begin{equation}
   \sigma_{1}(E,T=0)
   =\left\{\begin{array}{ll}
           \sigma_{1}^{c}  &  (E=E_{c}) \\
           0               &  (E\neq E_{c})
           \end{array}
      \right.
\label{ct0}
\end{equation}
here $\sigma_{1}^{c}$ is the universal conductance[8] at the mobility
edge $E_c$ of the quantum Hall liquid.  At finite
temperatures, inelastic scattering brings in another length scale,
$L_{in} \propto T^{-p/2}$, beyond which electronic wavefunction loses
coherence. This inelastic scattering length provides an effective
sample size at finite temperatures. The transport properties of the
system are now determined by the competition between  these two length scales
$L_{in}(T)$ and $\xi (E)$. For example, at extremely low temperature,
$L_{in}(T)$ is essentially infinite, $\xi (E)$ is the dominant length
scale, QHE is observed everywhere except at a set of discrete energies
$E_c$ near the center of LL.
With increasing $T$, $L_{in}(T)$ decreases and states with $\xi
(E)\simgt L_{in}(T)$ effectively become
delocalized, and the system behaves like a metal with the
disappearance of QHE in the regime $\xi
(E)\simgt L_{in}(T)$ .
So at finite temperatures, a finite band of delocalized states near the
critical energy is formed contributing to the longitudinal
conductivity. In the region far from the critical energy,
localization has no scaling behavior,  and,  at very low temperatures
variable-range
hopping transport dominates the dissipative conductivity
$\sigma_{xx}$.
Our scaling theory does not apply to the variable-range hopping regime.

 Near the conductance peak({\it i.e.} in between QH steps), $E_F$ is
close to $E_c$, and,
 $\rho_{xy}$ and $\rho_{xx}$
are both scaling functions of a single scaling variable which can
be written as:
\begin{equation}
 v=(\frac{L_{in}(T)}{\xi (E)})^{1/\nu }\propto |E-E_{c}|T^{-p/2\nu }
\end{equation}
with the experimental critical exponent being  expressed as $\kappa =p/2\nu
$[3].
In this context, we propose, following Ref.6, that the
universal conductance near the
scaling regime in one Landau band at finite temperatures should also be
a scaling function of $v$, {\em i.e.}
\begin{equation}
    \sigma_{1}(E,T)=\sigma_{1}^{c}S_{1}(v)
\end{equation}
where we can choose a simple form for $S_1(v)$ as
\begin{equation}
S_{1}(v)=e^{-v^{2}}=exp\left[-\frac{(E-E_{c})^{2}}{\gamma_{0}T^{2\kappa}}\right]=exp\left[-\frac{(E-E_{c})^{2}}{\Gamma^{2}}\right]
\end{equation}
where $\gamma_{0}$ is a non-universal constant.
This simple model not only satisfies the condition given by Eq.(\ref{ct0})
but also takes into account the delocalized band around $E_c$ at finite
temperatures. The band width is given by $\Gamma
=\sqrt{\gamma_{0}}T^{\kappa }$ which is consistent with the experiment
of  Ref.3 where it is found that $\Delta B \propto T^{\kappa }$.
We have tried a number of alternate reasonable forms for $S_1(v)$ such
as $S_1(v)=e^{-|v|^{\nu}}$, and, $S_1(v)$ given by a constant within a cut
off. We get essentially the same numerical results as the ones
presented in this paper with these alternate forms of $S_1(v)$.

   Even though our model ignores the contribution of the
variable-range hopping conduction, in the intermediate temperature range
where thermally activated contribution from the scaling conductivity
dominates, our model should provide a good approximation.
Combining Eqs.(2)-(6) we find that
the contribution of a single Landau band to the thermally
activated longitudinal conductivity can be written as (set $k_{B}=1$):
\begin{equation}
  \sigma_{xx}=\sigma_{1}^{c}\int_{-\infty}^{\infty}\frac{dE}{T}
\frac{e^{(E-E_{F})/T}}{(1+e^{(E-E_{F})/T})^{2}}exp\left[-\frac{(E-E_{c})^{2}}{\Gamma^{2}}\right]
\end{equation}
When the Fermi energy is in the quantum Hall plateau region, we have
to take into account contribution from both Landau bands below and
above the Fermi energy because of the electron-hole symmetry. Assuming the
Fermi energy $E_{F}$ is in between two critical energies, and
defining the activation energy gap $\Delta =|E_{F}-E_{c}|$, we can write
$\sigma_{xx}$ as
\begin{equation}
  \sigma_{xx}=2\sigma_{1}^{c}\int_{-\infty}^{\infty}\frac{dE}{T}
\frac{e^{E/T}}{(1+e^{E/T})^{2}}exp\left[-\frac{(E-\Delta
)^{2}}{\Gamma^{2}}\right]
\label{ctt}
\end{equation}
As argued in Ref.6, at low enough temperatures where $k_{B}T<<\Delta $,
Fermi-Dirac distribution function can be replaced by a Boltzmann
distribution function,
immediately producing Eq.(\ref{s2}) for $\sigma_{xx}$ with
\begin{equation}
  S_{2}(\frac{T}{\Gamma })=\frac{\sqrt{\pi\gamma_{0}}}{T^{1-\kappa
}}exp\left[\frac{\gamma_{0}}{4T^{2(1-\kappa)}}\right]
\end{equation}
which is a strong function of temperature, and is manifestly non-universal.
We note that, according to Eq.(9), even though $2\sigma_{1}^{c}$ itself
is universal,
the thermally activated $\sigma_{xx}$ does not really have a simple
exponentially activated form with a universal prefactor.
Also, one finds by putting $T=0$ in Eq.(9) that $S_2(0)$ is actually
violently divergent rendering invalid[7] the simple argument for
universality made in Ref.6. Our numerical results show that Eq.(9) is
not really quantitatively valid in any temperature regime for the
available experimental data.

     In Fig.1, we show the best-fit of our model (Eq.(\ref{ctt})) with
the actual
experimental data taken from  Ref.4. In the best-fit analysis, we
use the  Fermi-Dirac distribution function. We assume the universal
conductance to be[8] $2\sigma_{1}^{c}=e^{2}/h$. The energy gap is
obtained for even integer filling as
$\Delta =\frac{1}{2}(\hbar\omega_{c}-g^{*}\mu_{B}B)$,
where $g^{*}(=0.44)$ is the unrenormalized $g$-factor. We therefore need to
do a 2-parameter fitting to obtain $\kappa $ and $\gamma_{0}$.
As shown in Fig.1, the critical exponent $\kappa $ slowly varies
around $\kappa\simeq 0.35$ for different Landau levels. We also find
that $\gamma_{0}$ changes non-universally for different Landau levels
(as shown in Table I) and, in fact, the calculated $\Gamma $ from our
fits is smaller than, but
comparable to,  $\Delta $ for higher Landau bands. As we can see in
the plot, except for the low temperature variable-range hopping
regime where the scaling analysis does not apply,
higher LL, in general, provide better
fitting in the scaling region. This is expected from the fact that the
relative scaling regime ($\Gamma /\Delta $) is much larger  for higher Landau
levels. The slight disagreement for  lower Landau
levels implies
that our scaling model does not describe very well the tail of the
scaling conductivity which tends to dominate the conductivity at
the intermediate temperature range. This is again expected because one
does not expect scaling to hold very far from $E_c$.
We define a new variable
\begin{equation}
     \Delta '=\frac{\partial ln\sigma_{xx}}{\partial (1/T)}.
\end{equation}
If $\sigma_{xx}$ has the exponentially activated form({\em i.e.}
$\sigma_{xx}\propto e^{-\Delta /T}$), $\Delta '$ should be a constant
and should have the value of the activation energy $\Delta$({\it i.e.}
the energy gap). In the
inset of Fig.1, we show the calculated temperature dependence of $\Delta '$ for
$\nu =10$ for our model. We can see that in the experimental scaling
temperature range, {\it i.e.} away from the variable range hopping
regime,
 $\Delta '$ is very far from being a constant. This
supports our conclusion that $\sigma_{xx}$ cannot really be expressed
as a simple exponentially
activated form, and, a naive extrapolation to $1/T=0$ may be dangerous
in some situations. We emphasize, however, that an extrapolation (as is
clear from Fig.1 and 2) provides the correct order of magnitude of
$\sigma_1^c$.

We have carried out a similar fitting analysis of the scaling theory
to the recent data of Ref.5. These results are shown in Fig.2 (and,
Table I). Again, the scaling fit works very well and the values of
fitting parameter $\kappa$ and $\gamma_0$ are reasonable. The behavior
for $\Delta '$ in Ref.5 is similar to that in Ref.4.

In summary, our proposed scaling model works  well for higher Landau
levels because the scaling region (characterized by $\Gamma/\Delta$)
is large for higher LL.
The scaling theory
is not expected to be a very accurate model for
the universal conductance at the tail of the scaling regime, which
makes our model quantitatively less accurate for lower LL where
$\Gamma/\Delta$ is small.
Our calculated value of the exponent $\kappa\approx 0.3-0.5$ is
actually very consistent with the experimentally determined $\kappa$
from the conductivity peak measurements[3,9].
Our analysis establishes that scaling concepts in two dimensional
strong field localization apply not only
to the $\sigma_{xx}$ peaks in between the quantum Hall steps, but also
to the activated conduction regime for $\sigma_{xx}$ minima at the
quantum Hall steps provided the variable hopping range contribution is
negligible.
\vspace{0.3cm}

The authors thank R.N.Bhatt, R.G.Clark, S.Kivelson, A.H.MacDonald, and
D.C.Tsui for helpful discussions.
 This work is supported by the National Science Foundation.

\pagebreak

\begin{center}
{\Large \bf References}
\end{center}

\begin{flushleft}

[1] {\em The Quantum Hall Effect}, edited by R.E.Prange
and S.M.Girvin (Springer-Verlag, New York, 1990).\\

[2] D.Z.Liu and S. Das Sarma, Mod.Phys.Lett. B {\bf 7}, 449(1993) and
references therein.\\

[3] H.P.Wei, D.C.Tsui, M.Paalanen, and A.M.M.Pruisken,
Phys.Rev.Lett. {\bf 61}, 1294 (1988).\\

[4] R.G.Clark, Physica Scripta, {\bf T39}, 45 (1991).\\

[5] Y.Katayama, D.C.Tsui, and M.Shayegan, preprint(1993).\\

[6] D.H.Lee, S.Kivelson, and S.C.Zhang, Phys.Rev.Lett. {\bf 68}, 2386
(1992).\\

[7] R.N.Bhatt, N.Read, and B.Huckestein, unpublished; S.Kivelson,
private communication.\\

[8] Y.Huo, R.E.Hetzel, and R.N.Bhatt, Phys.Rev.Lett. {\bf 70},
481(1993).\\

[9] S.Koch, R.J.Haug, K. von Klitzing, and K.Ploog, Phys.Rev.Lett.
{\bf 67}, 883(1991).

\end{flushleft}

\pagebreak

\begin{center}
{\Large \bf Figure and Table Caption}
\end{center}

\begin{flushleft}
{\bf FIG.1} Temperature activated longitudinal conductance (in the
units of $e^{2}/h$). $(\Diamond )$s are the experimental data
(The sample carrier density is $n_{s}=2.20\times
10^{11}cm^{-2}$)
presented in Ref. 4.
 Solid lines are the best-fit of the data by our scaling
model with the best-fit values of the exponent $\kappa$.
 Inset: Temperature dependence of $\Delta '$
 (for $\nu=10$), the dashed line shows
 corresponding energy gap $\Delta$.

{\bf FIG.2} The same as in Fig.1 for the data in Ref.5. The inset
shows $\Delta '$(for $\nu=5$).

\vspace{1cm}
{\bf TABLE I} Delocalized state band widths $\Gamma $ for different
Landau levels for the data of Ref.4 and 5(the lowest two rows).

\end{flushleft}
\begin{tabular}{||c|c|c|c|c||}  \hline\hline
$\nu $  &  $\gamma_{0} $  &  $\Gamma (T=1K) $ & $\Delta (K)$ & $\Gamma
/\Delta (\% )$ \\ \hline
  2  &  9.60  & 3.10  & 45.60 &  6.8 \\ \hline
  4  &  7.69  & 2.77  & 22.80 & 12.1 \\ \hline
  6  &  7.18  & 2.68  & 15.20 & 17.6 \\ \hline
  8  &  4.94  & 2.22  & 11.40 & 19.5 \\ \hline
 10  &  3.86  & 1.96  &  9.12 & 21.5 \\ \hline
  3  &  0.66  & 0.81  &  2.86 & 28.4 \\ \hline
  5  &  0.71  & 0.84  &  1.56 & 54.0 \\ \hline
\hline
\end{tabular}

\end{document}